\documentstyle[eqsecnum,aps,epsf]{revtex} % PH. REV. FINAL FORMAT STYLE
\begin{document}

\twocolumn[\hsize\textwidth\columnwidth\hsize\csname@twocolumnfalse\endcsname

\title{Free expansion of attractive and repulsive Bose-Einstein condensed
vortex states}

\author{Sadhan K. Adhikari}
\address{Instituto de F\'{\i}sica Te\'orica, Universidade Estadual
Paulista, 01.405-900 S\~ao Paulo, S\~ao Paulo, Brazil\\}

\date{\today}
\maketitle
\begin{abstract}

Free expansion of attractive and repulsive Bose-Einstein condensed vortex
states formed in an axially symmetric trap is investigated using the
numerical solution of the time-dependent Gross-Pitaevskii equation.  In a
repulsive condensate the vortex-core radius is much smaller than the
radial root-mean-square (rms) radius, which makes the
experimental observation of the vortex core difficult.  The opposite is
found to be true in an attractive condensate which makes it a better
candidate for experimental observation. Also, in all cases the ratio of
vortex-core radius to radial rms radius increases as the angular momentum
of the vortex increases. This makes the vortex states with higher angular
momenta more suitable for experimental confirmation.

{\bf PACS Number(s):  03.75.Fi}

\end{abstract}

\vskip1.5pc]
 \newpage

\section{Introduction}
 
Since the successful detection \cite{1} of Bose-Einstein condensates
(BEC) in dilute bosonic atoms employing magnetic trap at ultra-low
temperature, one problem of extreme interest is the formation of vortex
states in an axially symmetric trap. The recent experimental detection
\cite{exp} of vortex states in the condensate has intensified this
endeavor. There have been many theoretical studies on different aspects of
BEC \cite{2,3,11} and specially,  on the vortex states in axially
symmetric traps \cite{2a,2b,2c,2d,2e,2f,2g,2h}.  

The study of the free expansion of a vortex state is of utmost importance
as such an expansion can aid in its experimental confirmation \cite{2b}.
The specific behavior of the time evolution of vortex core and the radius
of the atomic core after the removal of the trap could be fundamental in
the identification of vortex states.  However, a previous theoretical
study on the free expansion of vortex states was limited to condensates
with repulsive interatomic interaction \cite{2b}. In this work we extend
this effort to condensates with attractive interatomic interaction and
compare critically the expansion dynamics with that for repulsive
interaction.  This is of extreme experimental relevance as by changing the
external electromagnetic field now it is possible to change the effective
interatomic interaction in a controlled way from repulsive to attractive
\cite{yy1}.

We base our study on the numerical solution \cite{3,2a,2b,2h} of the
nonlinear time-dependent mean-field Gross-Pitaevskii (GP)  equation
\cite{8} with an axially symmetric trap, which should provide a faithful
description of the formation and evolution of vortex states \cite{2b,2h}.

The study of superfluid properties of BEC is of great interest to both
theoreticians and experimentalists \cite{2c,2d}.  Quantized vortex state
in BEC is intimately connected to the existence of superfluidity. Such
vortices are expected in superfluid He$_{\mbox{II}}$. However,
due to strong interaction between the helium atoms there is no reliable
mean-field description and a controlled theoretical study of the dynamics
is not possible.

On the other hand, a weakly interacting trapped BEC is well-described
by the mean-field GP equation that is known to admit vortex solutions for
a trap with cylindrical symmetry \cite{2a}, which can be studied
numerically. In contrast to superfluid He$_{\mbox{II}}$, the condensed
vortex state of trapped BEC is an extremely dilute 
quantum fluid and permits a mean-field description.
Moreover, vortex states in BEC have been observed
experimentally \cite{exp}. Many different techniques for creating
vortex states in BEC have been suggested \cite{2d}, e.g., stirring the
BEC with an external laser \cite{2c}, spontaneous vortex formation in
evaporative cooling \cite{2f}, and rotation of axially symmetric trap
\cite{2g}.

It is possible to have dynamically stable vortex BEC states with low
quanta of rotational excitation or angular momentum $L$ per particle for
both attractive \cite{2a,2g} and repulsive \cite{2h} atomic interactions.
However, in the attractive case such states exist for the number of atoms
below a critical number which increases with $L$.  In the absence of
vortex ($L=0$), the stable condensate in a cylindrical trap has a
cylindrical shape.  Such a BEC has the largest density on the axis of the
trap.  In the presence of vortex motion the region of largest density of
the BEC with nonzero $L$ is pushed away from the central axial region and
the atoms have more space to stabilize. The vortex state of the condensate
in a cylindrical trap has the shape of a hollow cylinder with zero density
on the axis of symmetry.

For vortex states with low angular quanta $L$ and large number of atoms,
the radius of the hollow part is much smaller than the radius of the
condensate in the radial direction, which makes the experimental
confirmation of vortex states difficult. It has been argued and
established in repulsive condensates with $L=1$ that after a free
expansion, the hollow core of the vortex may expand faster than the radius
of the condensate and thus make the internal hollow region detectable
experimentally confirming a vortex state \cite{2b}.

In our study of free expansion of vortex states we find that the
attractive condensates allow vortex states with a different nature from
those in repulsive condensates. For a fixed $L$, the ratio of vortex core
radius to radial root-mean-square (rms) radius of an attractive condensate
is always larger than the same ratio for a repulsive condensate, which may
make attractive condensates more appealing for generating, detecting and
studying vortex states.  We also find that for a fixed $L$, upon free
expansion the above ratio increases for repulsive condensates, whereas it
decreases for attractive condensates remaining, however, always larger
than the corresponding ratio for repulsive condensates.  In all cases the
ratio increases with $L$.

The present study is performed with the direct numerical solution of the
time-dependent GP equation with an axially symmetric trap. In the
time-evolution of the GP equation the radial and axial variables are dealt
with in two independent steps. In each step the GP equation is solved by
discretization with the   Crank-Nicholson  rule 
complimented by the known boundary
conditions \cite{2h,koo}. We find that this time-dependent
approach leads to good convergence.

In Sec. II we describe briefly the time-dependent 
GP equation with  vortex states and a numerical method for
its solution. In
In Sec. III we report the numerical results of the present investigation
about the free expansion of vortex states 
and finally, in
Sec. IV we give a summary of our investigation.

\section{Nonlinear Gross-Pitaevskii Equation}

The deatil of the calculational procedure based on the GP equation has
been elaborated elsewhere \cite{2h} and we present a summary of our
account here.
At zero temperature, the time-dependent Bose-Einstein condensate wave
function $\Psi({\bf r};\tau)$ at position ${\bf r}$ and time $\tau $ may
be described by the following  mean-field nonlinear GP equation
\cite{11,8}
\begin{eqnarray}\label{a} \biggr[ -\frac{\hbar^2}{2m}\nabla^2
&+& V({\bf r})  
+ gN|\Psi({\bf
r};\tau)|^2\nonumber \\
&-&i\hbar\frac{\partial
}{\partial \tau} \biggr]\Psi({\bf r};\tau)=0.   \end{eqnarray} Here $m$
is
the mass and  $N$ the number of atoms in the
condensate, 
 $g=4\pi \hbar^2 a/m $ the strength of interatomic interaction, with
$a$ the atomic scattering length. 
The trap potential with cylindrical symmetry may be written as  $  V({\bf
r}) =\frac{1}{2}m \omega ^2(r^2+\lambda^2 z^2)$ where 
 $\omega$ is the angular frequency
in the radial direction $r$ and 
$\lambda \omega$ that in  the
axial direction $z$. We are using the cylindrical
coordinate system ${\bf r}\equiv (r,\theta,z)$ with $\theta$ the azimuthal 
angle.
The normalization condition of the wave
function is
$ \int d{\bf r} |\Psi({\bf r};t)|^2 = 1. $

The GP
equation (\ref{a}) can 
accommodate
quantized vortex
states with rotational motion of the condensate around the $z$ axis.
In such a vortex the atoms flow with
tangential velocity $L\hbar/(mr)$ such that each atom has quantized 
angular momentum
$L\hbar$ along $z$ axis. This corresponds to the  angular dependence 
\begin{equation}\label{ang}
\Psi({\bf r};\tau)=\psi(r,z;\tau)\exp (iL\theta)
\end{equation}
 of the wave
function.

Substituting Eq. (\ref{ang}) into
Eq.  (\ref{a}), and transforming to 
dimensionless variables
defined by $x =\sqrt 2 r/l$,  $y=\sqrt 2 z/l$,   $t=\tau \omega, $ $l\equiv \sqrt {\hbar/(m\omega)}$, 
and  
\begin{equation}\label{wf}
\phi(x,y;t)\equiv 
\frac{ \varphi(x,y;t)}{x} =  \sqrt{\frac{l^3}{\sqrt 8}}\psi(r,z;\tau),
\end{equation} 
we get 
\begin{eqnarray}\label{d}
\biggr[ -\frac{\partial^2}{\partial
x^2}&+&\frac{1}{x}\frac{\partial}{\partial x} -\frac{\partial^2}{\partial
y^2}
+\frac{L^2}{x^2}
+\frac{1}{4}\left(x^2+\lambda^2 y^2-\frac{4}{x^2}\right) \nonumber \\
&+& 8 \sqrt 2 \pi   n\left|\frac {\varphi({x,y};t)}{x}\right|^2 
-i\frac{\partial
}{\partial t} \biggr]\varphi({ x,y};t)=0, 
\end{eqnarray}
where
$ n =   N a /l.$ 
The normalization condition  of the wave
function become \begin{equation}\label{5} {2\pi} \int_0 ^\infty
dx \int _{-\infty}^\infty dy|\varphi(x,y;t)|
^2 x^{-1}=1.  \end{equation}

For a stationary solution the time dependence of the wave function is
given by $\varphi(x,y;t) = \exp(-i\mu t ) \varphi(x,y)$
where $\mu$ is the chemical potential of the condensate in units of
$\hbar\omega$. If we use this
form of the wave function in Eq. (\ref{d}), we obtain the following
stationary nonlinear time-independent GP equation \cite{8}: 
\begin{eqnarray}\label{dx}
\biggr[ -\frac{\partial^2}{\partial
x^2}&+&\frac{1}{x}\frac{\partial}{\partial x} -\frac{\partial^2}{\partial
y^2}+\frac{L^2}{x^2}
+\frac{1}{4}\left(x^2+\lambda^2 y^2-\frac{4}{x^2}\right) \nonumber \\
&+&8\sqrt 2 \pi n\left|\frac {\varphi({x,y})}{x}\right|^2 
-\mu  \biggr]\varphi({ x,y})=0. 
\end{eqnarray}
The 
rms
radii of the condensate in the radial and axial directions  are defined,
respectively, by 
\begin{equation}\label{7a}  (x_{\mbox{rms}})^2
= 2\pi \int_0
^\infty dx \int _{-\infty}^\infty dy x |\varphi(x,y;t)| ^2,
\end{equation}
and 
\begin{equation}\label{7b}  (y_{\mbox{rms}})^2
= 2\pi
\int_0
^\infty dx \int _{-\infty}^\infty dy x^{-1}y^2 |\varphi(x,y;t)| ^2.
\end{equation}

We solve the GP equation (\ref{d}) using a time-iteration method
elaborated in Refs. \cite{2h,koo}.  The full GP Hamiltonian is
conveniently
broken
into two parts $H_x$ and $H_y$ $-$ the first containing the $x$-dependent
terms and the second containing the $y$-dependent terms with the nonlinear
interaction term divided equally into two parts. The GP equations along 
the $x$ and $y$ directions  are
defined on a two-dimensional set of grid points $N_x \times N_y$ using the 
Crank-Nicholson discretization method. The resultant tridiagonal equations
along $x$ and $y$ directions are solved alternately  by the
Gaussian elimination method along the $x$ and $y$
directions \cite{koo}. Effectively, each time
iteration of the GP equation is broken up into two parts  $-$ first using
$H_x$
and then $H_y$. 
For a small time step $\Delta$ the error involved in this
break-up procedure along $x$ and $y$ directions is quadratic in $\Delta$
and hence can be neglected. This scheme is repeated for about 500 
time
iterations to yield the final solution.

The initial solution
for the iterative scheme is taken to be the analytic solution of the
corresponding free harmonic oscillator setting the nonlinear term to zero
in the GP equation (\ref{dx}).  At each iteration the strength of the
nonlinear term
is increased by a small amount so that after the final iteration the full
strength is attained and the required solution of the GP equation 
obtained. 
The free expansion of the vortex states is studied essentially using 
the above iterative scheme by setting the trapping potential to zero in
the GP
equation after the final solution for the vortex state is obtained.

The typical value of the space step used for discretization along $x$ and
$y$ directions is 0.02 and that for time step 
$\Delta$ is 0.05. For small
nonlinearity the largest values of $x$ and $y$ are
  $x_{\mbox{max}} = 8$, $|y|_{\mbox{max}} =
8$. However, for stronger nonlinearity, larger values of  $x_{\mbox{max}}$
and $|y|_{\mbox{max}}$ (up to 20) are employed.

\section{Numerical Result}

We present results for the free expansion of vortex states using the
numerical solution of the time-dependent GP equation for attractive and
repulsive interatomic interactions for different $L$. In a recent
study
of repulsive vortex states with $L=1$, Dalfovo et al. \cite{2b} pointed
out that the
small vortex core radius of these condensates make the vortex states hard
to detect experimentally. They show that after a free
expansion, the
vortex-core radius increases faster than the radial rms radius, which
makes the experimental observation of the vortex core easier. Here we
complement the above study by including the attractive condensates as well
as vortex states with $L> 1$. The interesting features of the results that
we emphasize here are quite independent of the value of axial to radial
trap frequency 
$\lambda$, and in
this study we present results for $\lambda=\sqrt 8$ only.

In Figs. 1(a) and (b) we plot the profile of the section of
the wave function $|\phi(x,0;t)|\equiv |\varphi(x,0;t)|/x$ upon free
expansion as a function of $x$  at different times $0<t<3$ for
$n=-1.5$ (attractive case) and $n=12$
(repulsive case), respectively,  $-$ the trap being removed
at time $t=0$. In the attractive case, the minimum value of $n$ allowed
for $L=1$ is $-1.63$ \cite{2h}.  A vortex-core radius $x_c$ is
conveniently
defined
to be the smaller value of $x$ where the central core density
$|\phi(x,0;t)|^2$ attains $e^{-1}$ times the peak value in Figs. 1(a) and
(b). After free expansion, both the core radius $x_c$ and radial rms
radius of the condensate increase. The relative increase of these two
radii is crucial in the experimental observation of the vortex states. For
this a careful study of the evolution of the different radii is important.
In Fig. 2 we present the time evolution of $x_{\mbox{rms}}$ and
$y_{\mbox{rms}}$,
where
we plot $x_{\mbox{rms}}(t)/x_{\mbox{rms}}(0)$ and
$y_{\mbox{rms}}(t)/y_{\mbox{rms}}(0)$ vs. $t$
for $L=1$ and $n=-1.5, 0, 12$. Both ratios 
increase with time in all cases. In the attractive case the radius in the
radial direction increases faster, the opposite being true in the
repulsive case. The result for $n=0$ always stays between the attractive
and repulsive cases. 

Next we study the evolution of the ratio $x_c/x_{\mbox{rms}}$, always
less
than unity, with time $t$. The larger this ratio, the larger is the
vortex-core radius compared to the radial rms radius and the easier could
be the experimental detection of the vortex state. In Fig. 3 we plot
$x_c/x_{\mbox{rms}}$ vs. time $t$ for $L=1$ and 4 for several $n$ including
the
attractive (negative $n$), repulsive (positive $n$) and noninteracting
($n=0$) cases. For fixed $n$ and $t$, this ratio increases with $L$;  and
for fixed $L$ and $t$, the ratio increases as $n$ decreases from positive
to negative values (passes from repulsive to attractive).  This means that
for attractive interaction the vortex state has a larger vortex-core
radius compared to the radial rms radius and hence could be easily
detected. From Fig. 3 we find that at $t=0$ the numerical 
value of this ratio  for attractive condensate could be twice as large 
as its value for repulsive condensate. 
However, there is an interesting difference
between the time evolution of this ratio for a fixed $L$ for attractive
and repulsive interactions. This ratio decreases with time for attractive
interaction and increases with time for repulsive interaction. Hence there
is no specific advantage in allowing an attractive  vortex to expand for
the sake of detection. It will be easier to detect it before expansion,
when the vortex core is more pronounced. 
For the
noninteracting case the above  ratio remains constant at all times $-$
$\sim
0.28$ for $L=1$ and $\sim 0.60$ for $L=4$.  Hence, although the
possibility of detection  of the repulsive vortex state increases upon
free expansion,
the opposite is true for attractive interaction. These features of the
attractive vortex states upon expansion are the most important results of
this investigation clearly displayed in Fig. 3.

We also studied the expansion of vortex states of  higher angular momenta 
in detail.
The general features of these states are similar and we present the
results 
for $L=4$ in the following.
In Figs. 4 (a) $-$ (d) we plot the profile of the complete wave function
$|\phi(x,y;t)|$ vs. $x$ and $y$ 
for $n=-4.3$ for $t=0,1,2$ and 3.  In this attractive case the critical
$n$ for
collapse is $-4.4$ \cite{2h}. In Figs. 5 and 6 we plot the same for
$n=4.3$
and 16, respectively. 
At all times   the wave function
is zero on the $y$ axis and is
peaked at some finite $x$.  Figures 4 $-$ 6 clearly show the time
evolution of the full condensate in these cases. We find from these
figures that at a particular  time, the vortex core radius $x_c$ is
approximately the same in the three cases. However, the radial rms radius 
$x_{\mbox{rms}}$ increases as $n$ increases from negative to positive
values and consequently, the ratio $x_c/x_{\mbox{rms}}$ increases as we
pass from Fig. 4 to 5 and then to 6 (from attractive to repulsive).  
The evolution of the ratio $x_c/x_{\mbox{rms}}$ upon free expansion for
$L=4$ and
different $n$ is shown in Fig. 3. The results for the 
$L=4$ case is qualitatively similar to the $L=1$ case and the discussion and
comments of the $L=1$ case apply here.

An interesting feature of the freely expanding condensate emerges from the
present study. The vortex core radius $x_c$ and the radial rms radius 
$x_{\mbox{rms}}$ play important roles in this game. The
condensed vortex state has  the form of a hollow cylinder. Qualitatively, 
$x_{\mbox{rms}}$ is the radius of the cylinder and $x_c$ is the radius of
the
hollow. The hollow space is more prominent when  $x_c/x_{\mbox{rms}}$ is
larger.  A large value of $x_c/x_{\mbox{rms}}$ facilitates the experimental
confirmation of the vortex state.  
From Figs. 1, 2,  4, 5 and 6 we find that both  $x_c$ and $x_{\mbox{rms}}$
increase upon free expansion. However, for an
attractive  condensate $x_{\mbox{rms}}$ increases at a faster rate than
$x_c$
so that the above ratio decreases with time. The opposite is true for a 
repulsive condensate. For a noninteracting condensate both radii increase
at the
same rate and the ratio remains constant. The whole scenario  is physically
understandable. For an attractive condensed vortex state upon free
expansion, because of
interatomic attraction, the radius of the hollow $x_c$ can not increase
fast enough with time compared to the radius of the cylinder. The cylinder
as a whole swells at a faster rate than the
hollow  making the ratio 
$x_c/x_{\mbox{rms}}$ decrease with time. The opposite happens for a
repulsive 
condensate, where the atomic repulsion makes the hollow part swell at
a much faster rate. For a noninteracting condensate $x_c$ and
$x_{\mbox{rms}}$
increase
at the same rate leading to a constant $x_c/x_{\mbox{rms}}$.  In general
for a
fixed $L$,  $x_{\mbox{rms}}$ is much smaller for an attractive condensate,
which
makes the ratio  $x_c/x_{\mbox{rms}}$ much larger than the same for a
repulsive
condensate.

\section{Summary}

In this paper we present a numerical study of the time-dependent
Gross-Pitaevskii equation for an axially symmetric  trap to
obtain insight into the free expansion of  vortex states of BEC.  The
time-dependent GP equation is solved iteratively by discretization using a
two-step Crank-Nicholson scheme \cite{2h,koo}.  

The ratio of vortex core radius to radial rms radius $x_c/x_{\mbox{rms}}
(< 1)$ 
is found to play an
interesting role in the free expansion of condensed vortex states. 
The larger this ratio, more prominent is the vortex core and 
the more easier is the possibility of experimental
detection of vortex states. For stable vortex states of fixed angular
momentum per particle $L$, this ratio for attractive interaction is larger
than that for repulsive interaction. Upon free expansion of such states,
the ratio for attractive interaction decreases remaining, however,  always
larger
than that for repulsive interaction, which increases after expanion. The
ratio increases with $L$ in all cases. This makes the stable attractive
vortex
states before expansion more suitable for experimental detection.

The work is supported in part by the Conselho Nacional de Desenvolvimento
Cient\'\i fico e Tecnol\'ogico and Funda\c c\~ao de Amparo \`a Pesquisa do
Estado de S\~ao Paulo of Brazil.

\vskip 1cm

{\bf Figure Caption:}

1. Wave function $\phi (x, y=0) $ of  a stable vortex state with $L=1$
vs. $x$ for (a) $n = -1.5$, and (b) $n=12$.

2. The ratios $x_{\mbox {rms}}(t)/x_{\mbox {rms}}(0)$
and  $y_{\mbox {rms}}(t)/y_{\mbox {rms}}(0)$ vs. time $t$ 
of expanding vortex states with $L=1$ and $n=-1.5, 0, 12$.

3. The ratio $x_{c}(t)/x_{\mbox {rms}}(t)$
vs. time $t$ 
of expanding vortex states with $L=1$, $n=-1.5, 0, 12$
and $L=4$, $n=\pm 4.3, 0, 16$.

4. The wave functions $\phi(x,y)$ of a freely expanding vortex state 
vs. $x$ and $y$ at  times $t=0,1,2,3$ for $L=4$ and $n=-4.3$. 

5. Same as Fig. 4  for $n= 4.3$.

6. Same as Fig. 4  for $n= 16$.

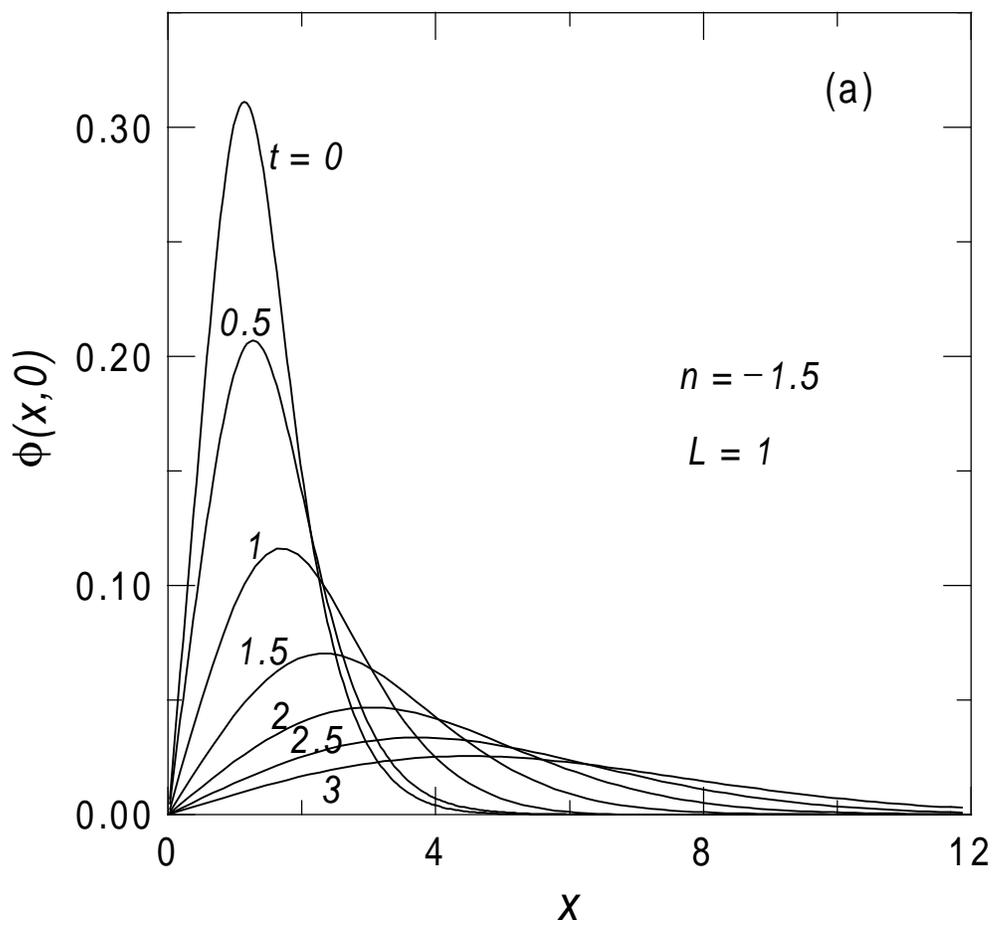
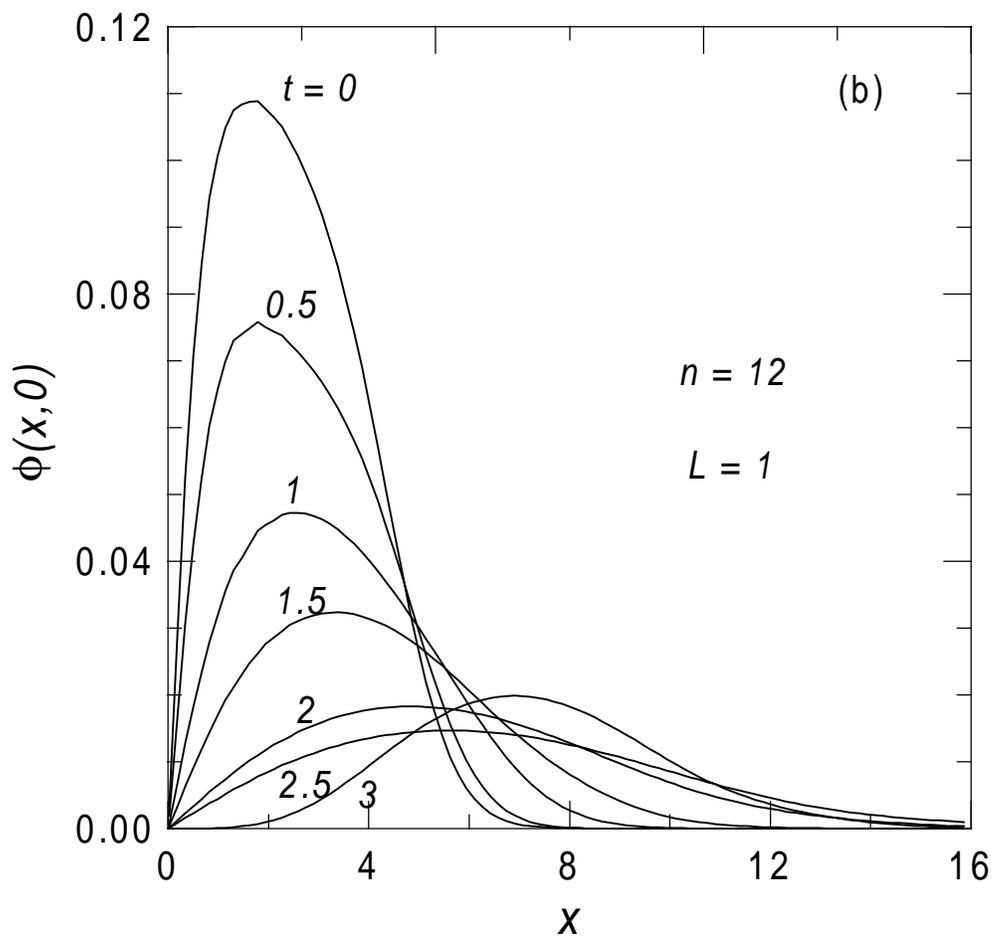
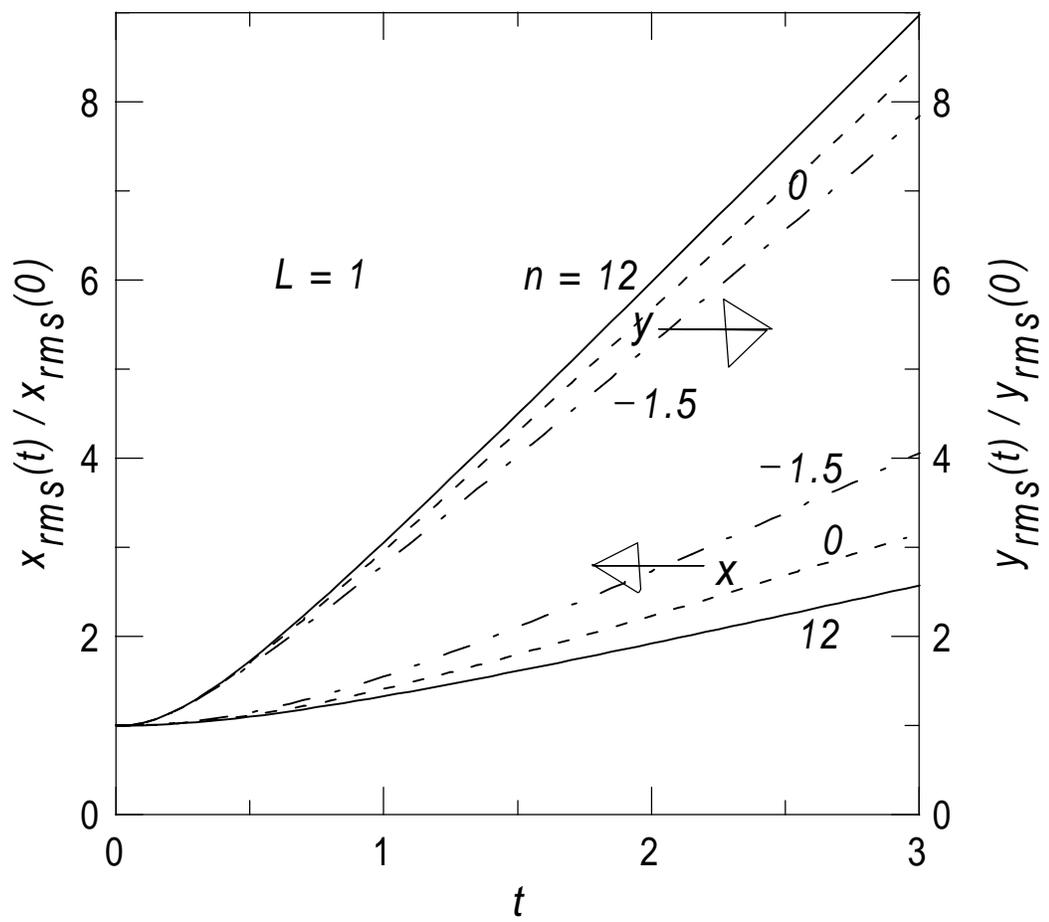
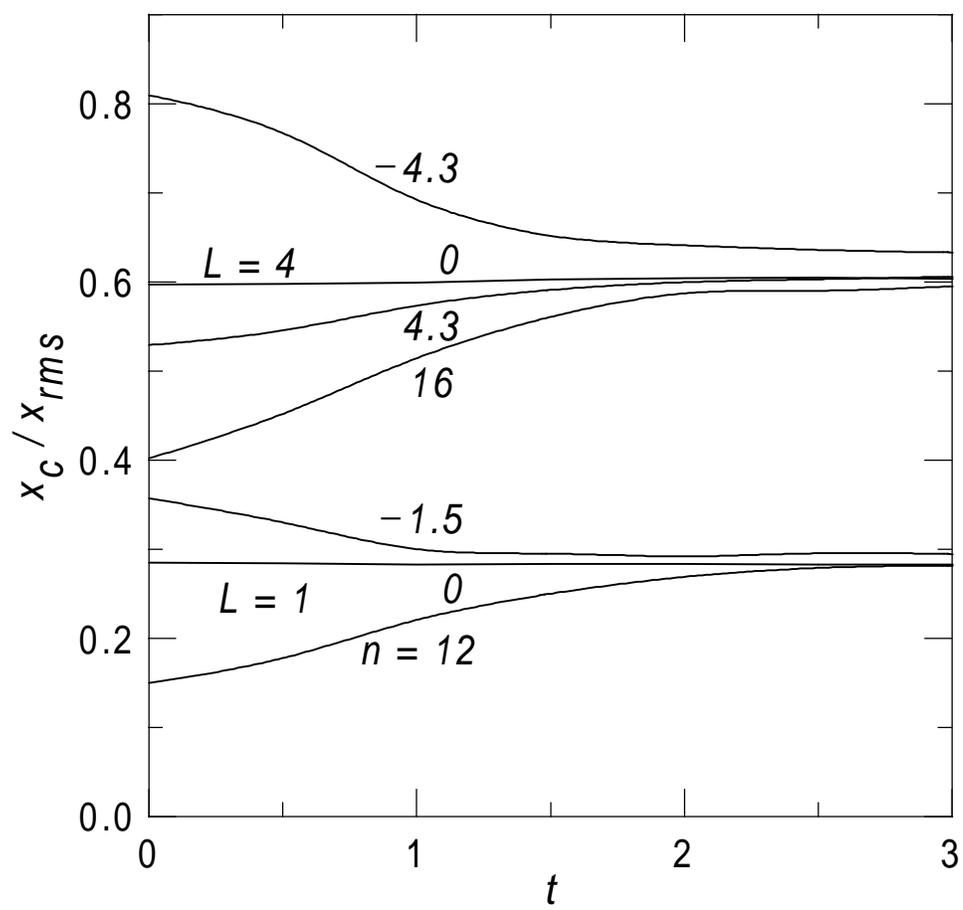
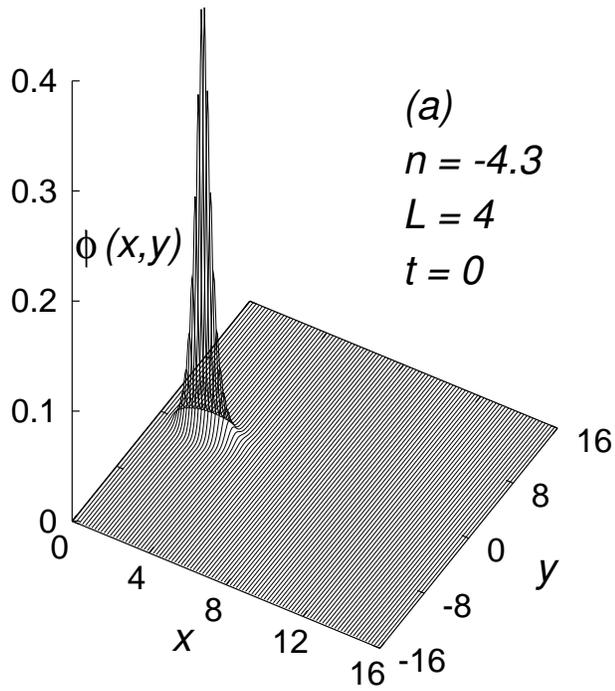
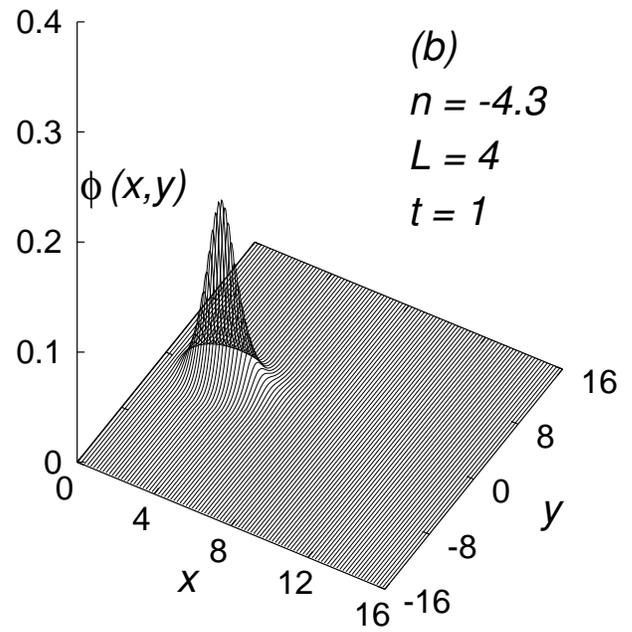
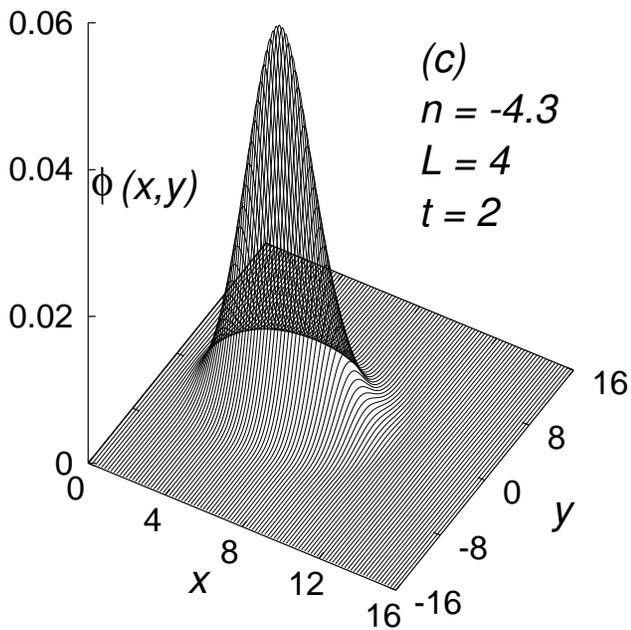
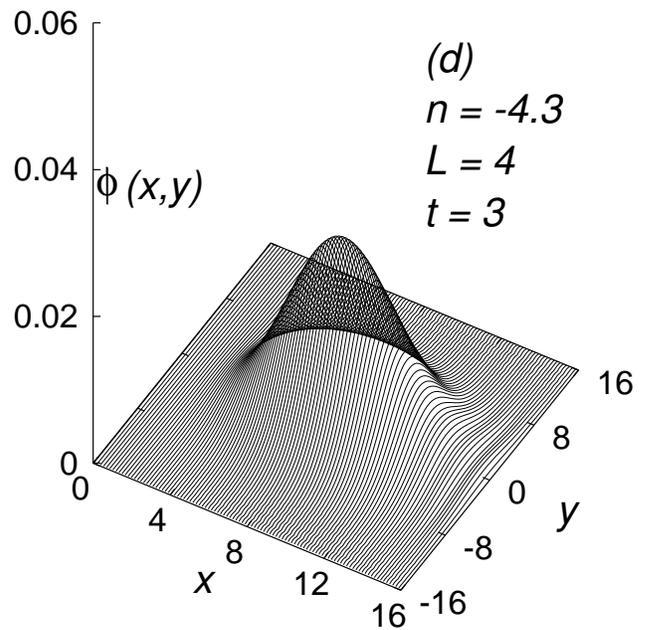
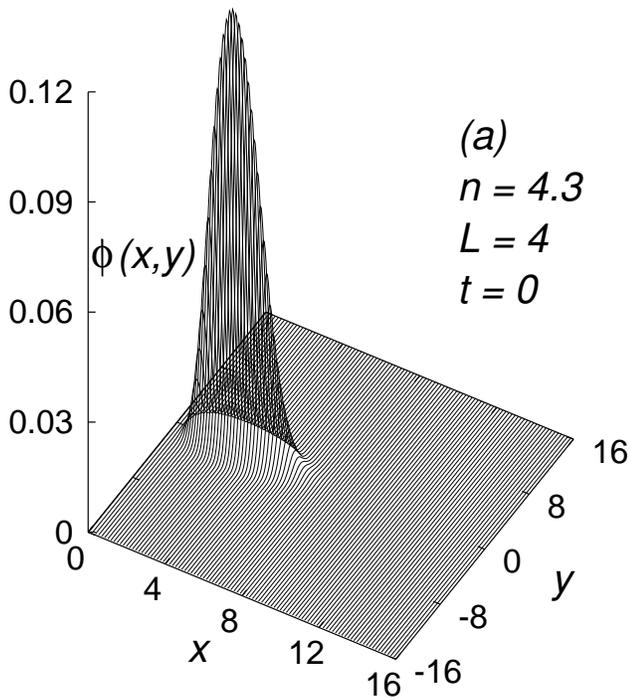
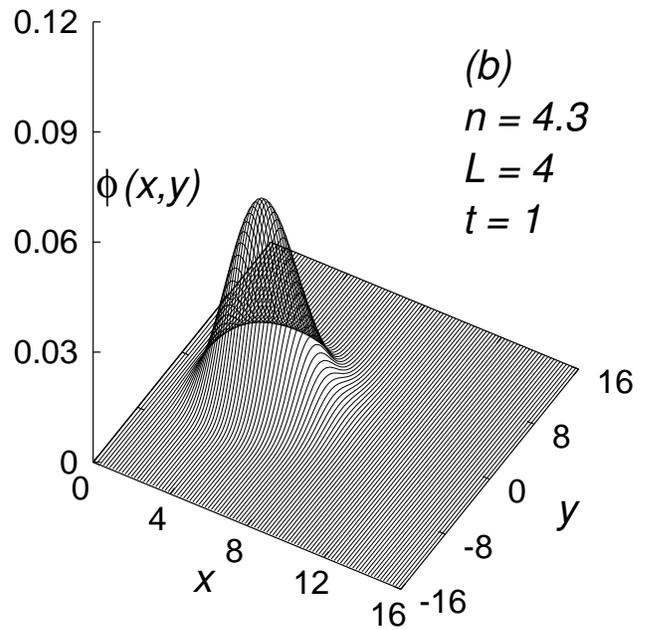
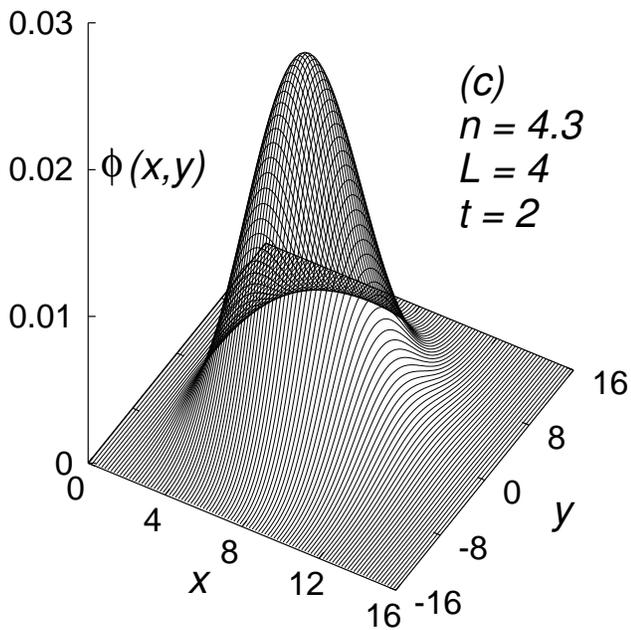
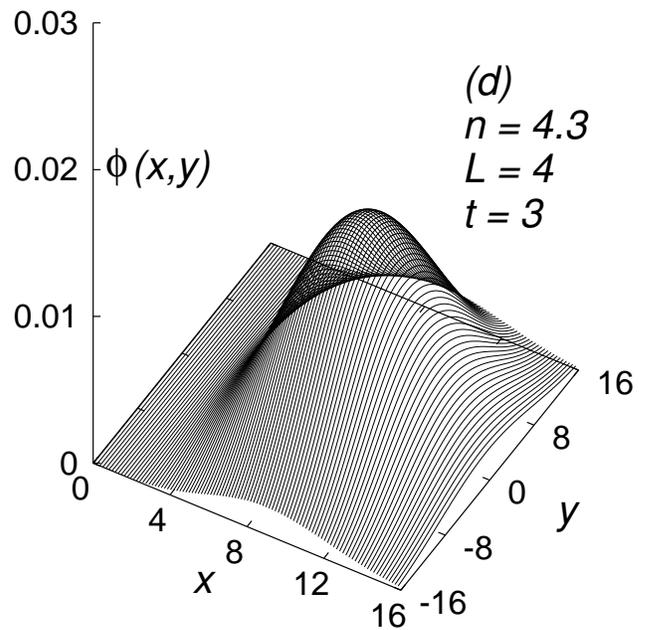
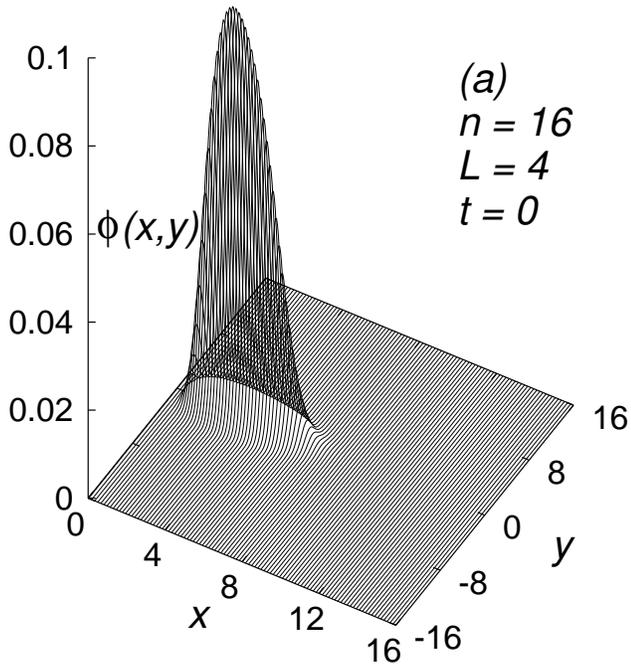
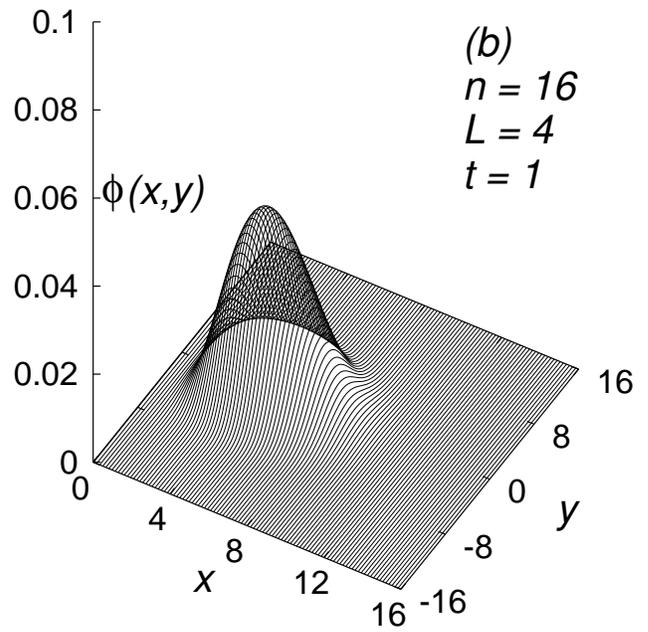
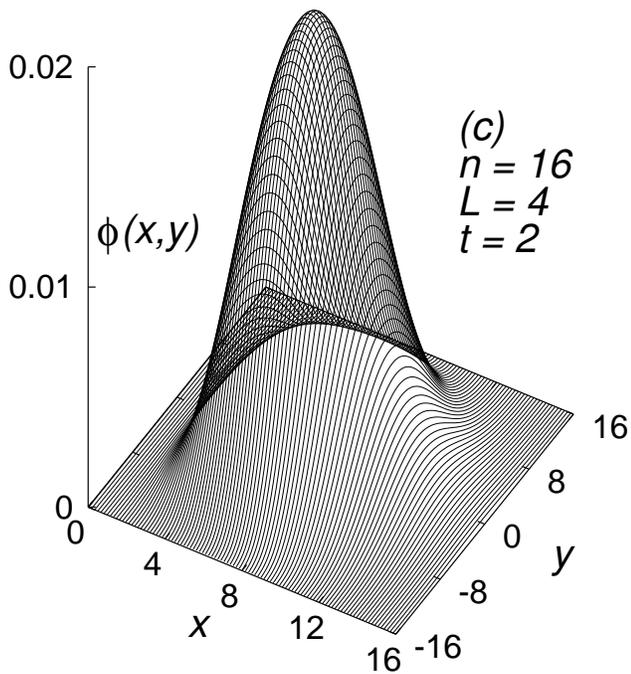
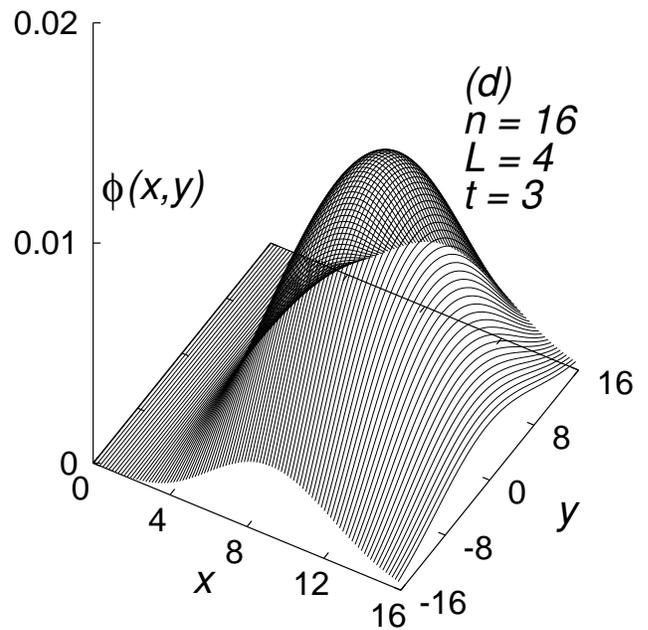
\end{document}